\documentclass[aps,reprint,prd,groupedaddress]{revtex4-2}
\usepackage{color}
\usepackage{float}
\usepackage{amsmath,amsfonts,amssymb,epsfig,graphicx}
\usepackage{slashed} 
\usepackage{hyperref}
\newcommand{\MGMCatNLO}{MadGraph5\_aMC@NLO} 
\newcommand{\pythia}{{\sc Pythia}}
\newcommand{\delphes}{{\sc Delphes}}
\newcommand{\fastjet}{{\sc FastJet}}

\newcommand{\pt}{\ensuremath{p_\mathrm{T}}}

\newcommand{\PZ}{\ensuremath{\mathrm{Z}}}
\newcommand{\PH}{\ensuremath{\mathrm{H}}}
\newcommand{\PW}{\ensuremath{\mathrm{W}}}

\newcommand{\mup}{\ensuremath{\mu^+}}
\newcommand{\mum}{\ensuremath{\mu^-}}
\newcommand{\elp}{\ensuremath{\mathrm{e}^+}}
\newcommand{\elm}{\ensuremath{\mathrm{e}^-}}
\newcommand{\nui}{\ensuremath{\nu_i}}
\newcommand{\nul}{\ensuremath{\nu_\ell}}

\newcommand{\nue}{\ensuremath{\nu_{\mathrm{e}}}}
\newcommand{\nuae}{\ensuremath{\bar{\nu}_{\mathrm{e}}}}
\newcommand{\num}{\ensuremath{\nu_{\mu}}}
\newcommand{\nuam}{\ensuremath{\bar{\nu}_{\mathrm{\mu}}}}
\newcommand{\mnu}{\sum_{i} m_{\nui}}

\newcommand{\eV}{\,\text{eV}}

\newcommand{\GeV}{\,\text{GeV}}
\def\lag{{\cal L}}
\newcommand{\confirm}[1]{{\color{black}#1}}
\newcommand{\mn}{{\rm M}_{\rm{N}}}

\newcommand{\fbinv}{\mbox{\ensuremath{~\mathrm{fb^{-1}}}}}

\graphicspath{{./pic/}}
\begin{document}

\title{The physics case for neutrino-neutrino collisions}

% \affiliation command applies to all authors since the last \affiliation command. % \affiliation can be followed by \email, \homepage, \thanks as well.
%\thanks{}

\author{Sitian \surname{Qian}}
\email[]{stqian@pku.edu.cn}

\author{Tianyi \surname{Yang}}
\email[]{tyyang99@pku.edu.cn}

\author{Sen \surname{Deng}}
%\email[]{sdeng@cern.ch}

\author{Jie \surname{Xiao}}
%\email[]{jiexiao@pku.edu.cn}

\author{Leyun \surname{Gao}}
%\email[]{seeson@pku.edu.cn}

\author{Andrew Michael \surname{Levin}}
\email[]{andrew.michael.levin@cern.ch}

\author{Qiang \surname{Li}}
\email[]{qliphy0@pku.edu.cn}

\affiliation{State Key Laboratory of Nuclear Physics and Technology, School of Physics, Peking University, Beijing, 100871, China}

\author{Meng \surname{Lu}}
%\email[]{meng.lu@cern.ch}

\author{Zhengyun \surname{You}}
%\email[]{youzhy5@mail.sysu.edu.cn}

\affiliation{School of Physics, Sun Yat-Sen University, Guangzhou 510275, China}

\begin{abstract}
Addressing the mass origin and properties of neutrinos is of strong interest to particle physics, baryogenesis and cosmology. Popular explanations involve physics beyond the standard model, for example, the dimension-5 Weinberg operator or heavy Majorana neutrinos arising from ``seesaw''  models. The current best direct limits on the electron neutrino mass, derived from nuclei beta decay or neutrinoless double beta decay processes, are at the sub-electronvolt level.  Here we propose a novel neutrino-neutrino collider where the neutrino beam is generated from TeV scale muon decays. Such collisions can happen between either neutrinos and anti-neutrinos, or neutrinos and neutrinos. We find that with a tiny integrated luminosity of about $10^{-5}$\fbinv we can already expect to observe direct neutrino anti-neutrino annihilation, $\nu\bar{\nu}\rightarrow {\rm Z}$, which also opens the door to explore neutrino related resonances $\nu\bar{\nu}\rightarrow {\rm X}$. The low luminosity requirement can accommodate a relatively large emittance muon beam. Such a device would also allow for probing heavy Majorana neutrino and effective Majorana neutrino mass through $\nu\nu\rightarrow\PH\PH$ to a competitive level, for both electron and muon types. 
\end{abstract}

\maketitle

\section{Introduction}
\label{introduction}

Neutrinos are among the most abundant and least understood of all particles in the Standard Model (SM) that make up our universe. The observation of neutrino oscillations confirms that at least two types of SM neutrinos have a tiny, but strictly nonzero, mass. The neutrino mass eigenstates $\nui$ (with $i=1,2,3$) are related to the weak eigenstates $\nul$ (with $\ell=\text{e},\mu,\tau$), via the neutrino mixing matrix:  $|\nui\rangle=\sum_{\ell} U_{i\ell}|\nul\rangle$. However, the absolute neutrino mass scale remains a mystery so far. 

Upper limits on each neutrino mass have been obtained with methods, including indirect cosmological constraints and direct measurements~\cite{PDG}. The final full-mission Planck measurements of the CMB anisotropies~\cite{Planck:2018vyg,Vagnozzi:2017ovm}, combined with baryon acoustic oscillation (BAO) measurements, constrain the neutrino mass~\cite{eBOSS:2020yzd} tightly to $\mnu < 0.12\,\eV$. However, these limits strongly rely on the underlying cosmological assumption. Direct measurement through fitting the shape of the beta spectrum, e.g., the Karlsruhe Tritium Neutrino (KATRIN) experiment~\cite{KATRIN:2021uub}, yields an upper limit $m_{\nuae} < 0.8\,\eV$ at the 90\% C.L., for the electron anti-neutrino $\nuae$, where $m^2_{\nuae}=\sum_i |U_{\mathrm{e}i}|^2 m_i^2$.  On the other hand, the direct mass limit on the electron neutrino $\nue$ and muon neutrinos are relatively much looser~\cite{PDG}. 

One of the simple formalism in which neutrino masses can arise is through a dimension-5 operator as shown by Weinberg~\cite{Weinberg:1979sa}, which extends the SM Lagrangian with 
\begin{align}
\lag_5 = \left( \ {\rm C}_5^{\ell\ell'}/\Lambda \right) \big[\Phi\!\cdot\! \overline{L}^c_{\ell }\big] \big[L_{\ell'}\!\!\cdot\!\Phi\big], 
\label{wbo}
\end{align}
where $\ell ,~\ell'$ are the flavors of the leptons, and can be electrons, muons or taus; $\Lambda$ is the relevant new physics scale; ${\rm C}_5^{\ell\ell'}$ is a flavor-dependent Wilson coefficient; $L_\ell^T=(\nu_\ell,\ell)$ is the left-handed lepton doublet; and $\Phi$ is the SM Higgs doublet with a vacuum expectation value $v=\sqrt{2}\langle\Phi\rangle\approx246\GeV$. The Weinberg operator generates the Majorana neutrino masses as $\confirm{m_{\ell\ell} = C_5^{\ell\ell} v^2/\Lambda}=\left|\sum_i U_{\mathrm{\ell}i}^2 m_i\right|$, and introduces lepton number violation (LNV). The unitarity violation (UV) completion of the Weinberg operator can be realized in the context of ``seesaw'' models~\cite{Minkowski:1977sc,Yanagida:1979as,Yanagida:1980xy,GellMann:1980vs,Mohapatra:1979ia}, assuming the existence of hypothetical heavy states, for example the heavy Majorana neutrino (HMN) in the type-I seesaw model. HMNs
can only couple to the SM through mixing with SM neutrinos, which is characterized by the
mixing element, $V_{\ell N}$, between an SM neutrino in its left-handed interaction state and a heavy 
 Majorana neutrino in its mass eigenstate. 
 
 Searches for neutrinoless double beta decay ($0\nu\beta\beta$) in the decays of heavy nuclei have placed strong limit, $m_{\beta\beta}<0.036-0.156$ eV at the 90\% C.L.~\cite{GERDA:2020xhi,KamLAND-Zen:2022tow}. On the other hand, neutrino oscillation~\cite{PDG} together with the KATRIN direct measurement~\cite{KATRIN:2021uub} also set stringent limit on $m_{\ell\ell}$, for all three generations of leptons. However, we do not yet know how neutrino masses are generated, and it is possible that there are multiple contributions at different scales to neutrino masses. If there are large cancellations between the Weinberg operator and other sources of neutrino masses, one would anticipate a signal search for neutrinoless double beta decay ($0\nu\mu\mu$).

The Weinberg operator induces couplings between neutrino and Higgs fields and thus $\nu \nu \rightarrow \PH\PH$ scattering. Inspired by the recently proposed neutrino lepton collider~\cite{Yang:2022qgs}, we would like to examine a novel method to probe neutrino mass through colliding collimated high energy neutrino beams. Such a device would allow for probing $\nu\nu\PH\PH$ couplings and thus effective Majorana neutrino mass to a competitive accuracy, for both electron and muon types. Above all, neutrino and anti-neutrino collision itself is a very interesting process to be explored, which can also open the door to probe neutrino related resonance or new physics beyond the SM. 

Finally we would like to inform readers that neutrino-neutrino collisions can also appear within the muon-muon collisions (see e.g. ref.~\cite{Han:2020uid}) at the same interaction point, however, this is out of the scope of this paper focusing on pure neutrino-neutrino collisions. Moreover, if the primary muon beams are allowed to collider, elastic scattering of $\mup\mum\rightarrow \mup\mum$, annihilation of $\mup\mum$ or $e^+e^-$, and muon beam induced background will all possibly bring huge backgrounds.

\section{Experimental Setup\label{sec:beam}}

As introduced in Refs.~\cite{King:1999kx, Yang:2022qgs}, collimated beams of neutrinos can be produced from a long straight section in the muon collider ring. For a muon beam with energy of 1.5 to 5 TeV, the circumference is from 4.5 to 10 km~\cite{Accettura:2023ked}, and the decayed neutrinos’ energy are spread from zero to their parental energy value~\cite{Yang:2022qgs}.  An illustration of the proposed neutrino beam can refer to Ref.~\cite{Yang:2022qgs}. While neutrino-neutrino collisions can be realised with two of such beams, a detailed configuration of the detector to be put in between the two neutrino beams is to be defined, with potential radiation hazards to be mitigated~\cite{King:1999kx}.

Similarly as discussed in Ref.~\cite{Yang:2022qgs}, the instantaneous luminosity of a neutrino collider would be limited by two main factors: 1) the intensity of the neutrino beam compared with the incoming muon beam is suppressed by roughly the fraction of the collider ring circumference occupied by the production straight section~\cite{King:1999kx}, 2) the neutrino beam spread, which may be modulated to be more focused.

In more details, by using the formula for the instantaneous luminosity,
\begin{align}
  {\cal L} = {N_{\rm beam 1} N_{\rm beam 2} \over 4 \pi \sigma_x \sigma_y} f_{\rm rep},
\end{align}
where $f_{\rm rep}$ is the rate of collisions and is typically 100 kHz (40 MHz) for lepton colliders (hadron colliders), and $N_{\rm beam 1,2}$ are the number of particles in each bunch which can be taken as $\sim 10^{11}\text{--}10^{12}$~\cite{FCC:2018evy}, $\sigma_x$ and $\sigma_y$ are the beam sizes. Take the LHC as an example, with $f_{\rm rep}=40$\,MHz, $\sigma_{x,y}=16$ microns, and $N_{\rm beam 1,2}=10^{11}$, one can get $ {\cal L}=10^{34}$ cm$^{-2}$s$^{-1}$. As for TeV muon colliders~\cite{Bossi:2020yne}, with $f_{\rm rep}=100$\,KHz, $\sigma_{x,y}\lesssim 10$ microns, and $N_{\rm beam 1,2}=10^{12}$, then $ {\cal L}=10^{33}\text{--}10^{34}$ cm$^{-2}$s$^{-1}$. 
For muon collider, $f_{\rm rep}$ can be lower as 15 Hz, however, taking into account the turn frequency, the effective number can still be 100 kHz~\footnote{https://online.kitp.ucsb.edu/online/muoncollider-m23/}.
As for the neutrino-neutrino collisions discussed above, there are further suppression factors from linear over arc ratio ($L^2_l/L^2_c\sim 1/100$) with the exact value depending on the realistic design as shown in Ref.~\cite{Yang:2022qgs}, and the neutrino beam spread which can be around 1000 microns for $L_l\sim$ 10 to 100 meters. Taking all these into account, a realistic instantaneous luminosity for neutrino-neutrino collisions can reach around $ {\cal L}=10^{28}$ cm$^{-2}$s$^{-1}$ level. 

\begin{figure}
    \centering
    \includegraphics[width=.9\columnwidth]{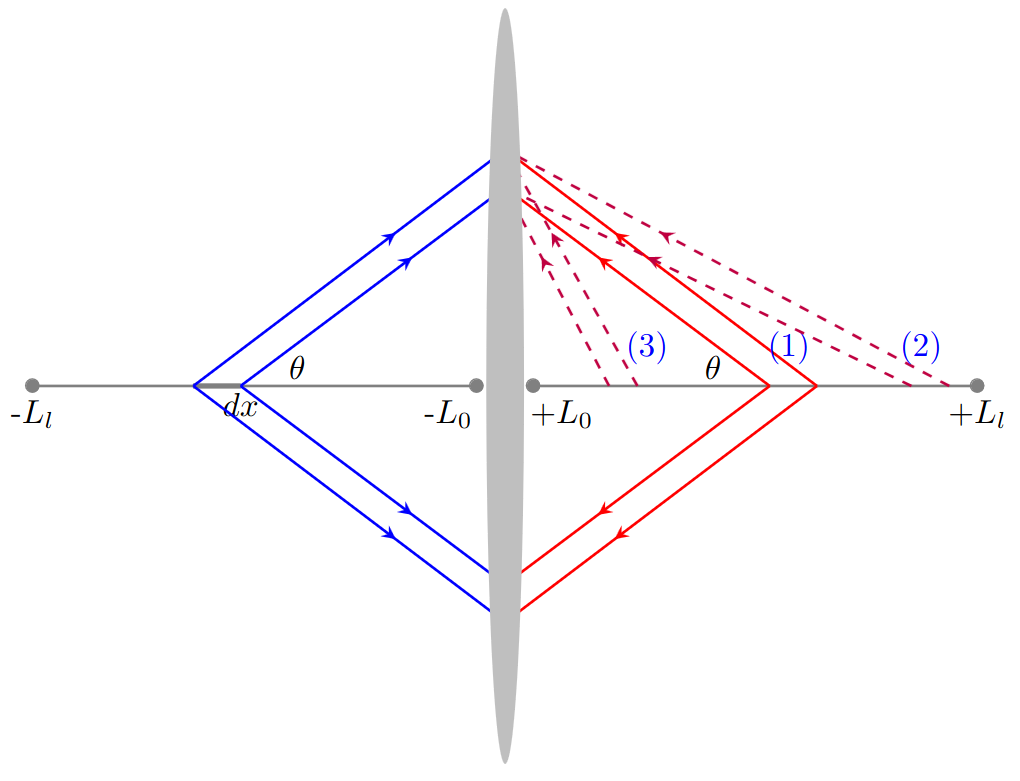}
    \caption{The neutrino-neutrino collision profile, where muon beams flying the linear path (symbolized by $L_l$ to $L_0$, and $L_0$ is a cut-off parameter defined by the muon beam size) radiate neutrinos approximately along a cone with polar angle as $\theta$. More explanations can be found in the texts.}
    \label{fig:lumi}
\end{figure}

To be more practical, we consider the neutrino-neutrino collision profile as shown in Fig.~\ref{fig:lumi},  where muon beams flying the linear path (symbolized by $L_l$ to $L_0$, where $L_0$ is a cut-off parameter defined by the muon beam size
%is the minimal length between muon beam and neutrinos' interaction point
) radiate neutrinos approximately along a cone with polar angle as $\theta\sim {\rm M}_\mu/{\rm E}_\mu$, i.e., the muon mass and energy ratio . We then estimate the instantaneous luminosity for neutrino-neutrino collisions as below:
\begin{align}
  {\cal L} &= \frac{L^2_l}{L^2_c} \int^{L_l}_{L_0}\frac{N_{\rm beam 1}  N_{\rm beam 2} f_{\rm rep}}{L^2_l\times (4\times 2\pi x\tan^2\theta)}\times dx \nonumber\\
  &= \frac{L^2_l}{L^2_c}\frac{N_{\rm beam 1} N_{\rm beam 2} f_{\rm rep}}{8\pi L^2_l\tan^2\theta}\times \ln(L_l/L_0),
\end{align}
with $L_l\tan\theta\sim r_s$, and there appears as an enhanced factor of $\ln(L_l/L_0)/2\sim 2-5$, and thus can further increase the instantaneous luminosity for neutrino-neutrino collisions. Note $L_0$ is a cut-off parameter in above integration formula and defined by the muon beam size, which can be at the order of 1-10 cm and thus may relax the stringent requirement on beam cooling of the nominal muon collider being pursued.

Notice in the integration, dominating contributions comes from (1) where neutrinos are radiated along the direction of $\theta$, while the contributions from the off-axis regions (2,3) can be approximated with the energy spread function as shown in Ref.~\cite{Yang:2022qgs} to be included in \MGMCatNLO~\cite{Alwall:2014hca} for simulations. The above estimation should serve as a reasonable estimation, although it may work worse for off-shell production with dominating contributions from asymmetric beam scenarios, where detailed Monte-Carlo studies are necessary. 

Although the above estimated instantaneous luminosity for neutrino-neutrino collisions is a small number, however, to reach the discovery threshold of neutrino antineutrino annihilation process $\nue \nuae \rightarrow \PZ$ to be discussed below, a tiny integrated luminosity of about $10^{-5}$\fbinv~is needed, i.e., several days or weeks of data taking.

\section{Physics Potential}
\label{amm}

We start from neutrino and antineutrino collision. With TeV scale $\mup\rightarrow \elp\nue\nuam$ and $\mum\rightarrow  \elm\nuae\num$ beams from two sides, there appears the collisions 
\begin{align}
&\nue \nuae \rightarrow \PZ \rightarrow \mup\mum. \label{eq:p0}
\end{align}
To simulate this process, we implement the neutrino energy fraction function~\cite{Yang:2022qgs} from 200 GeV muon decay in \MGMCatNLO~\cite{Alwall:2014hca}. Notice again, due to the complicated cone-like radiated neutrino profile from muon decay, our estimation is currently based on an approximation as discussed above, and may underestimate the contributions from asymmetric neutrino energy beam scenarios, where detailed Monte-Carlo studies are necessary. 

\begin{figure}
    \centering
    \includegraphics[width=0.5\textwidth]{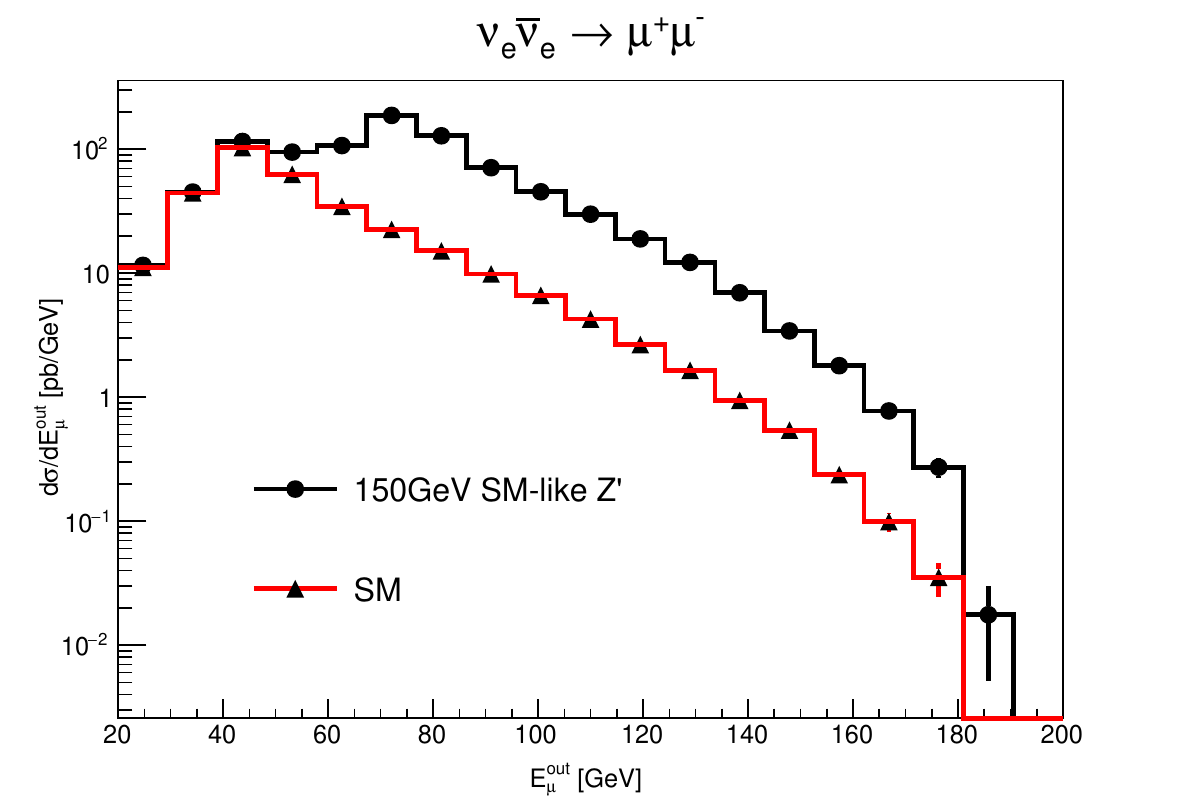}
    \caption{Outgoing muon energy distributions for neutrino antineutrino annihilation into Z and SM-like Z' bosons, with Z' mass set as 150 GeV with narrow width.}
    \label{fig:Emu}
\end{figure}

The cross section reads 320 pb, after requiring the final state muon to satisfy $\pt>20$\,GeV and $|\eta|<3.0$. As for the same process but with hadronic Z decay, the cross section will be around 5200 pb. Such a large cross section can compensate for the luminosity limitation as discussed above in Sec.~\ref{sec:beam}. For example, with a tiny integrated luminosity of about $10^{-5}$\fbinv, one can already expect to observe direct neutrino collisions through process.~(\ref{eq:p0}) and further can probe the $\PZ\nu\bar{\nu}$ couplings ~\cite{NuTeV:2001whx,Davidson:2001ji}, or search for possible $\nu\bar{\nu}$ resonance. Figure.~\ref{fig:Emu} shows the outgoing muon energy distributions for neutrino antineutrino annihilation into Z and SM-like Z' bosons, with Z' mass set as 150 GeV with narrow width.

On the other hand, we can also consider neutrino-neutrino collision, which is specifically sensitive to the Weinberg operator and Majorana neutrino mass. With 1 TeV $\mup\rightarrow  \elp\nue\nuam$ beams from two sides, some of the main physics processes can be shown as below (where we only consider $\nue$ for simplicity, while the results for $\num$  will be similar):
\begin{align}
&\nue \nue \rightarrow \PH\PH \label{eq:p1}\\
&\nue \nue \rightarrow \PZ\PZ\,, \PZ\PH \label{eq:p2}\\
&\nue \nue \rightarrow \nue \nue \PH, \label{eq:p3}\\
&\nue \nue \rightarrow \nue \nue \PZ\PZ\,, \nue\nue\PW\PW, \label{eq:p4} \\
&\nue \nue \rightarrow \nue \nue \PZ\PH\,, \nue \nue \PH\PH, \label{eq:p5} \\
&\nue \nue \rightarrow e^-e^- \PW^+ \PW^+, \label{eq:p6}
\end{align}

The first two are generated using the so called Phenomenological Type I Seesaw model~\cite{Fuks:2020att}, with scanned HMN mass from several to hundred TeV, and the HMN mixing element set as $V_{eN}=0.01$ by default (the cross sections are in proportional to $V^4$). The latter three arise from so called vector boson fusion processes in the SM. The cross sections numbers read:  11 fb for process.~(\ref{eq:p1}) and negligibly small for process.~(\ref{eq:p2}), with HMN mass $\mn=20$\,TeV;  133 fb for process.~(\ref{eq:p3}); 14 fb for process.~(\ref{eq:p4});  0.17fb for process.~(\ref{eq:p5}); 27 fb for process.~(\ref{eq:p6}) after requiring the final state electron to satisfy $\pt>10$\,GeV and $|\eta|<5.0$. 

\begin{figure}
    \centering
    \includegraphics[width=0.5\textwidth]{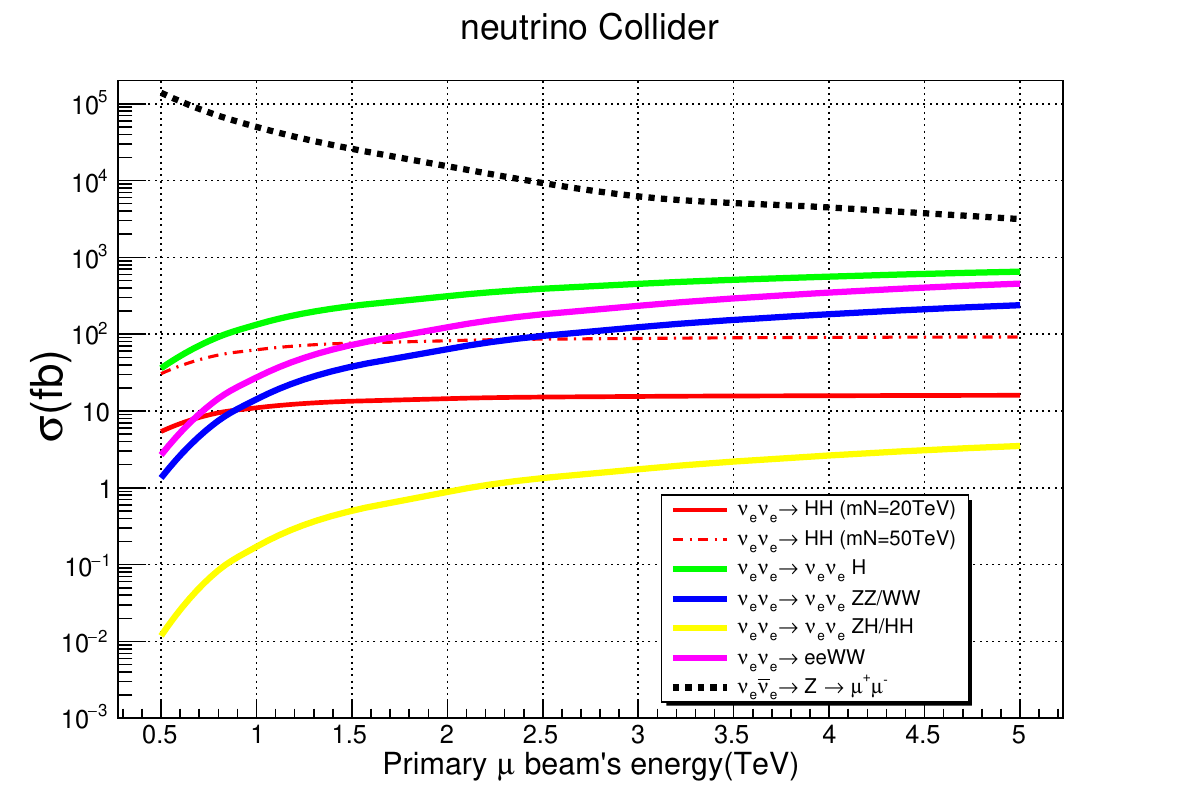}
    \caption{Cross section dependence on primary muon beam energy, for various neutrino-neutrino scattering processes~\ref{eq:p0}-\ref{eq:p6}. Final state leptons listed explicitly in the processes are required to satisfy $\pt>20$\,GeV and $|\eta|<3.0$.}
    \label{fig:xsec}
\end{figure}

The cross section variations on primary muon beam energy can be found in Fig.~\ref{fig:xsec}. One can see that cross sections of $\nue \nuae \rightarrow \PZ \rightarrow \mup\mum$ decrease with increasing beam energy in the range considered, while steadily increase for all other processes, due to different energy behaviour of the annihilation and t-channel processes.

\begin{figure}
    \centering
    \includegraphics[width=0.5\textwidth]{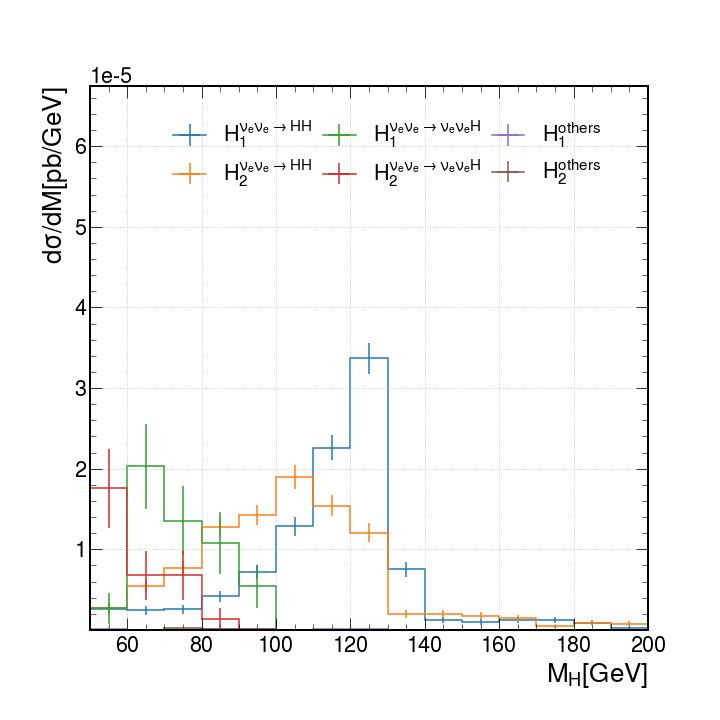}
    \caption{Mass distributions of the two reconstructed ``$\PH$ bosons'' from four b-tagged jets, for the signal process $\nue \nue \rightarrow \PH\PH$ and backgrounds, with neutrinos from 1 TeV muon beams' decay.  }
    \label{fig:mH}
\end{figure}

Both HMN signal ($\mn=20$\,TeV and $V_{eN}=0.01$) and background events are simulated with \MGMCatNLO, then showered and hadronized by \pythia8~\cite{Sjostrand:2014zea}. The final state jets are clustered using \fastjet~\cite{Cacciari:2011ma} with the VLC~\cite{Boronat_2018} algorithm at a fixed cone size of $R_{\rm jet}=0.7$. We used \delphes~\cite{deFavereau:2013fsa} version 3.5 to simulate detector effects with the default card for the muon collider detector~\cite{mucard}. We then require events include no well identified leptons while instead contain four b-tagged jets using the loose working point (90\% efficiency) with $\pt>30$~GeV and absolute pseudo-rapidity $|\eta|<2.5$. We cluster the selected four b-jets into two reconstructed ``$\PH$ bosons'' similar as we done for ``$\PZ$ bosons'' in Ref.~\cite{Yang:2021zak}. Figure.~\ref{fig:mH} shows the mass distributions of the two ``$\PH$ bosons'' reconstructed from the selected events. 

With 1\fbinv of data, by cutting on reconstructed ${\rm M}_{\PH}$, we are close to exclude $V_{eN}$ (and $V_{\mu N}$) $\gtrsim 0.01$ at $\mn=20$\,TeV, at 95\% C.L., which surpasses already current best limits from the CMS experiment~\cite{CMS:2022hvh} by two orders of magnitude. An interesting fact is that cross sections of $\nue \nue \rightarrow \PH\PH$ scale as $V^4_{eN}\mn^2$, thus this proposal can touch super heavy HMN region which is not possible in other experiments, and the resulted sensitivity on $V_{eN}$ scales inversely with $\sqrt{\mn}$. For example, for 1000 TeV HMN, the 95\% C.L exclusion limit can reach $V_{eN}\gtrsim 0.001$ with 1\fbinv of data, based on the same simulation study as described above. 

Accordingly, we are sensitive to the electron (muon) Majorana mass $m_{ee}\, (m_{\mu\mu})\sim V^2_{eN} (V^2_{\mu N}) \times\mn$ at the GeV level, exceeding already the sensitivity from previous collider experiments~\cite{CMS:2022hvh}, which can further be improved to reach MeV level with more data. 

%Note similarly one can exploit $\mum\rightarrow \elm\nuae\num$ beams. The same setup can thus be applied also to probe physics and masses related to neutrinos $\nue$, $\num$ and $\nuam$, which are loosely constrained to date. Finally, we mention that neutrino-neutrino collisions could also be sensitive to higher order neutrino EFT studies beyond the Weinberg operator~\cite{Cirigliano:2017djv}.

\section{Discussions}
\label{discussions}
Addressing the mass origin and properties of neutrinos is of strong interest to particle physics, baryogenesis and cosmology. Popular explanations involve physics beyond the standard model, for example, the dimension-5 Weinberg operator or heavy Majorana neutrinos arising from ``seesaw''  models. Current strongest direct search limits on heavy Majorana neutrino are below several TeV scale from the LHC experiment. Limits on electron neutrino mass are at sub-electronvolt level, derived from nuclei beta decay or neutrinoless double beta decay processes. Here we propose a novel idea on neutrino-neutrino collider where the neutrino beam is generated from TeV scale muon decays. With a tiny integrated luminosity of about $10^{-5}$\fbinv, one can already expect to observe direct neutrino anti-neutrino annihilation, which can accommodate a relatively large emittance muon beam. Such a device would also allow for probing neutrino related resonance (${\rm X}$) through $\nu\bar{\nu}\rightarrow {\rm X}$, or, heavy Majorana neutrino and effective Majorana neutrino mass through $\nu\nu\rightarrow\PH\PH$ to a competitive level, for various neutrino types $\nue$, $\nuae$, $\num$ and $\nuam$. It could also benefit searches for higher order neutrino EFT studies beyond the Weinberg operator.

\appendix
%\section{}
\begin{acknowledgments}
This work is supported in part by the National Natural Science Foundation of China under Grants No. 12150005, No. 12075004 and No. 12061141002, by MOST under grant No. 2018YFA0403900.
\end{acknowledgments}

% \begin{thebibliography}{10}
% \bibitem{PDG}
% P.~A.~Zyla \textit{et al.} [Particle Data Group],
% %``Review of Particle Physics,''
% PTEP \textbf{2020}, no.8, 083C01 (2020)
% doi:10.1093/ptep/ptaa104
% \bibitem{Cacciari:2008gp}
% M.~Cacciari, G.~P.~Salam and G.~Soyez,
% %``The anti-$k_t$ jet clustering algorithm,''
% JHEP \textbf{04}, 063 (2008)
% doi:10.1088/1126-6708/2008/04/063
% [arXiv:0802.1189 [hep-ph]].

% \bibitem{mucard} 
% https://github.com/delphes/delphes/blob/master/cards/delphes\_card\_MuonColliderDet.tcl

% \bibitem{CMS:2022zsu}
%  [CMS],
% %``Probing Majorana neutrinos and the Weinberg operator in the same-charge dimuon channel through vector boson fusion processes in proton-proton collisions at $\sqrt{s}=13~\mathrm{TeV}$ ,''
% CMS-PAS-EXO-21-003.

% \end{thebibliography}
% \bibliographystyle{ieeetr}
\bibliography{bibliography.bib}

\begin{thebibliography}{31}%
\makeatletter
\providecommand \@ifxundefined [1]{%
 \@ifx{#1\undefined}
}%
\providecommand \@ifnum [1]{%
 \ifnum #1\expandafter \@firstoftwo
 \else \expandafter \@secondoftwo
 \fi
}%
\providecommand \@ifx [1]{%
 \ifx #1\expandafter \@firstoftwo
 \else \expandafter \@secondoftwo
 \fi
}%
\providecommand \natexlab [1]{#1}%
\providecommand \enquote  [1]{``#1''}%
\providecommand \bibnamefont  [1]{#1}%
\providecommand \bibfnamefont [1]{#1}%
\providecommand \citenamefont [1]{#1}%
\providecommand \href@noop [0]{\@secondoftwo}%
\providecommand \href [0]{\begingroup \@sanitize@url \@href}%
\providecommand \@href[1]{\@@startlink{#1}\@@href}%
\providecommand \@@href[1]{\endgroup#1\@@endlink}%
\providecommand \@sanitize@url [0]{\catcode `\\12\catcode `\$12\catcode
  `\&12\catcode `\#12\catcode `\^12\catcode `\_12\catcode `\%12\relax}%
\providecommand \@@startlink[1]{}%
\providecommand \@@endlink[0]{}%
\providecommand \url  [0]{\begingroup\@sanitize@url \@url }%
\providecommand \@url [1]{\endgroup\@href {#1}{\urlprefix }}%
\providecommand \urlprefix  [0]{URL }%
\providecommand \Eprint [0]{\href }%
\providecommand \doibase [0]{https://doi.org/}%
\providecommand \selectlanguage [0]{\@gobble}%
\providecommand \bibinfo  [0]{\@secondoftwo}%
\providecommand \bibfield  [0]{\@secondoftwo}%
\providecommand \translation [1]{[#1]}%
\providecommand \BibitemOpen [0]{}%
\providecommand \bibitemStop [0]{}%
\providecommand \bibitemNoStop [0]{.\EOS\space}%
\providecommand \EOS [0]{\spacefactor3000\relax}%
\providecommand \BibitemShut  [1]{\csname bibitem#1\endcsname}%
\let\auto@bib@innerbib\@empty
%</preamble>
\bibitem [{\citenamefont {Workman}\ \emph {et~al.}(2022)\citenamefont {Workman}
  \emph {et~al.}}]{PDG}%
  \BibitemOpen
  \bibfield  {author} {\bibinfo {author} {\bibfnamefont {R.~L.}\ \bibnamefont
  {Workman}} \emph {et~al.} (\bibinfo {collaboration} {Particle Data Group}),\
  }\href {https://doi.org/10.1093/ptep/ptac097} {\bibfield  {journal} {\bibinfo
   {journal} {PTEP}\ }\textbf {\bibinfo {volume} {2022}},\ \bibinfo {pages}
  {083C01} (\bibinfo {year} {2022})}\BibitemShut {NoStop}%
\bibitem [{\citenamefont {Aghanim}\ \emph {et~al.}(2020)\citenamefont {Aghanim}
  \emph {et~al.}}]{Planck:2018vyg}%
  \BibitemOpen
  \bibfield  {author} {\bibinfo {author} {\bibfnamefont {N.}~\bibnamefont
  {Aghanim}} \emph {et~al.} (\bibinfo {collaboration} {Planck}),\ }\href
  {https://doi.org/10.1051/0004-6361/201833910} {\bibfield  {journal} {\bibinfo
   {journal} {Astron. Astrophys.}\ }\textbf {\bibinfo {volume} {641}},\
  \bibinfo {pages} {A6} (\bibinfo {year} {2020})},\ \bibinfo {note} {[Erratum:
  Astron.Astrophys. 652, C4 (2021)]},\ \Eprint
  {https://arxiv.org/abs/1807.06209} {arXiv:1807.06209 [astro-ph.CO]}
  \BibitemShut {NoStop}%
\bibitem [{\citenamefont {Vagnozzi}\ \emph {et~al.}(2017)\citenamefont
  {Vagnozzi}, \citenamefont {Giusarma}, \citenamefont {Mena}, \citenamefont
  {Freese}, \citenamefont {Gerbino}, \citenamefont {Ho},\ and\ \citenamefont
  {Lattanzi}}]{Vagnozzi:2017ovm}%
  \BibitemOpen
  \bibfield  {author} {\bibinfo {author} {\bibfnamefont {S.}~\bibnamefont
  {Vagnozzi}}, \bibinfo {author} {\bibfnamefont {E.}~\bibnamefont {Giusarma}},
  \bibinfo {author} {\bibfnamefont {O.}~\bibnamefont {Mena}}, \bibinfo {author}
  {\bibfnamefont {K.}~\bibnamefont {Freese}}, \bibinfo {author} {\bibfnamefont
  {M.}~\bibnamefont {Gerbino}}, \bibinfo {author} {\bibfnamefont
  {S.}~\bibnamefont {Ho}},\ and\ \bibinfo {author} {\bibfnamefont
  {M.}~\bibnamefont {Lattanzi}},\ }\href
  {https://doi.org/10.1103/PhysRevD.96.123503} {\bibfield  {journal} {\bibinfo
  {journal} {Phys. Rev. D}\ }\textbf {\bibinfo {volume} {96}},\ \bibinfo
  {pages} {123503} (\bibinfo {year} {2017})},\ \Eprint
  {https://arxiv.org/abs/1701.08172} {arXiv:1701.08172 [astro-ph.CO]}
  \BibitemShut {NoStop}%
\bibitem [{\citenamefont {Alam}\ \emph {et~al.}(2021)\citenamefont {Alam} \emph
  {et~al.}}]{eBOSS:2020yzd}%
  \BibitemOpen
  \bibfield  {author} {\bibinfo {author} {\bibfnamefont {S.}~\bibnamefont
  {Alam}} \emph {et~al.} (\bibinfo {collaboration} {eBOSS}),\ }\href
  {https://doi.org/10.1103/PhysRevD.103.083533} {\bibfield  {journal} {\bibinfo
   {journal} {Phys. Rev. D}\ }\textbf {\bibinfo {volume} {103}},\ \bibinfo
  {pages} {083533} (\bibinfo {year} {2021})},\ \Eprint
  {https://arxiv.org/abs/2007.08991} {arXiv:2007.08991 [astro-ph.CO]}
  \BibitemShut {NoStop}%
\bibitem [{\citenamefont {Aker}\ \emph {et~al.}(2022)\citenamefont {Aker} \emph
  {et~al.}}]{KATRIN:2021uub}%
  \BibitemOpen
  \bibfield  {author} {\bibinfo {author} {\bibfnamefont {M.}~\bibnamefont
  {Aker}} \emph {et~al.} (\bibinfo {collaboration} {KATRIN}),\ }\href
  {https://doi.org/10.1038/s41567-021-01463-1} {\bibfield  {journal} {\bibinfo
  {journal} {Nature Phys.}\ }\textbf {\bibinfo {volume} {18}},\ \bibinfo
  {pages} {160} (\bibinfo {year} {2022})},\ \Eprint
  {https://arxiv.org/abs/2105.08533} {arXiv:2105.08533 [hep-ex]} \BibitemShut
  {NoStop}%
\bibitem [{\citenamefont {Weinberg}(1979)}]{Weinberg:1979sa}%
  \BibitemOpen
  \bibfield  {author} {\bibinfo {author} {\bibfnamefont {S.}~\bibnamefont
  {Weinberg}},\ }\href {https://doi.org/10.1103/PhysRevLett.43.1566} {\bibfield
   {journal} {\bibinfo  {journal} {Phys. Rev. Lett.}\ }\textbf {\bibinfo
  {volume} {43}},\ \bibinfo {pages} {1566} (\bibinfo {year}
  {1979})}\BibitemShut {NoStop}%
\bibitem [{\citenamefont {Minkowski}(1977)}]{Minkowski:1977sc}%
  \BibitemOpen
  \bibfield  {author} {\bibinfo {author} {\bibfnamefont {P.}~\bibnamefont
  {Minkowski}},\ }\href {https://doi.org/10.1016/0370-2693(77)90435-X}
  {\bibfield  {journal} {\bibinfo  {journal} {Phys. Lett. B}\ }\textbf
  {\bibinfo {volume} {67}},\ \bibinfo {pages} {421} (\bibinfo {year}
  {1977})}\BibitemShut {NoStop}%
\bibitem [{\citenamefont {Yanagida}(1979)}]{Yanagida:1979as}%
  \BibitemOpen
  \bibfield  {author} {\bibinfo {author} {\bibfnamefont {T.}~\bibnamefont
  {Yanagida}},\ }\href@noop {} {\bibfield  {journal} {\bibinfo  {journal}
  {Conf. Proc. C}\ }\textbf {\bibinfo {volume} {7902131}},\ \bibinfo {pages}
  {95} (\bibinfo {year} {1979})}\BibitemShut {NoStop}%
\bibitem [{\citenamefont {Yanagida}(1980)}]{Yanagida:1980xy}%
  \BibitemOpen
  \bibfield  {author} {\bibinfo {author} {\bibfnamefont {T.}~\bibnamefont
  {Yanagida}},\ }\href {https://doi.org/10.1143/PTP.64.1103} {\bibfield
  {journal} {\bibinfo  {journal} {Prog. Theor. Phys.}\ }\textbf {\bibinfo
  {volume} {64}},\ \bibinfo {pages} {1103} (\bibinfo {year}
  {1980})}\BibitemShut {NoStop}%
\bibitem [{\citenamefont {Gell-Mann}\ \emph {et~al.}(1979)\citenamefont
  {Gell-Mann}, \citenamefont {Ramond},\ and\ \citenamefont
  {Slansky}}]{GellMann:1980vs}%
  \BibitemOpen
  \bibfield  {author} {\bibinfo {author} {\bibfnamefont {M.}~\bibnamefont
  {Gell-Mann}}, \bibinfo {author} {\bibfnamefont {P.}~\bibnamefont {Ramond}},\
  and\ \bibinfo {author} {\bibfnamefont {R.}~\bibnamefont {Slansky}},\
  }\href@noop {} {\bibfield  {journal} {\bibinfo  {journal} {Conf. Proc. C}\
  }\textbf {\bibinfo {volume} {790927}},\ \bibinfo {pages} {315} (\bibinfo
  {year} {1979})},\ \Eprint {https://arxiv.org/abs/1306.4669} {arXiv:1306.4669
  [hep-th]} \BibitemShut {NoStop}%
\bibitem [{\citenamefont {Mohapatra}\ and\ \citenamefont
  {Senjanovic}(1980)}]{Mohapatra:1979ia}%
  \BibitemOpen
  \bibfield  {author} {\bibinfo {author} {\bibfnamefont {R.~N.}\ \bibnamefont
  {Mohapatra}}\ and\ \bibinfo {author} {\bibfnamefont {G.}~\bibnamefont
  {Senjanovic}},\ }\href {https://doi.org/10.1103/PhysRevLett.44.912}
  {\bibfield  {journal} {\bibinfo  {journal} {Phys. Rev. Lett.}\ }\textbf
  {\bibinfo {volume} {44}},\ \bibinfo {pages} {912} (\bibinfo {year}
  {1980})}\BibitemShut {NoStop}%
\bibitem [{\citenamefont {Agostini}\ \emph {et~al.}(2020)\citenamefont
  {Agostini} \emph {et~al.}}]{GERDA:2020xhi}%
  \BibitemOpen
  \bibfield  {author} {\bibinfo {author} {\bibfnamefont {M.}~\bibnamefont
  {Agostini}} \emph {et~al.} (\bibinfo {collaboration} {GERDA}),\ }\href
  {https://doi.org/10.1103/PhysRevLett.125.252502} {\bibfield  {journal}
  {\bibinfo  {journal} {Phys. Rev. Lett.}\ }\textbf {\bibinfo {volume} {125}},\
  \bibinfo {pages} {252502} (\bibinfo {year} {2020})},\ \Eprint
  {https://arxiv.org/abs/2009.06079} {arXiv:2009.06079 [nucl-ex]} \BibitemShut
  {NoStop}%
\bibitem [{\citenamefont {Abe}\ \emph {et~al.}(2023)\citenamefont {Abe} \emph
  {et~al.}}]{KamLAND-Zen:2022tow}%
  \BibitemOpen
  \bibfield  {author} {\bibinfo {author} {\bibfnamefont {S.}~\bibnamefont
  {Abe}} \emph {et~al.} (\bibinfo {collaboration} {KamLAND-Zen}),\ }\href
  {https://doi.org/10.1103/PhysRevLett.130.051801} {\bibfield  {journal}
  {\bibinfo  {journal} {Phys. Rev. Lett.}\ }\textbf {\bibinfo {volume} {130}},\
  \bibinfo {pages} {051801} (\bibinfo {year} {2023})},\ \Eprint
  {https://arxiv.org/abs/2203.02139} {arXiv:2203.02139 [hep-ex]} \BibitemShut
  {NoStop}%
\bibitem [{\citenamefont {Yang}\ \emph {et~al.}(2022)\citenamefont {Yang},
  \citenamefont {Qian}, \citenamefont {Deng}, \citenamefont {Xiao},
  \citenamefont {Gao}, \citenamefont {Levin}, \citenamefont {Li}, \citenamefont
  {Lu},\ and\ \citenamefont {You}}]{Yang:2022qgs}%
  \BibitemOpen
  \bibfield  {author} {\bibinfo {author} {\bibfnamefont {T.}~\bibnamefont
  {Yang}}, \bibinfo {author} {\bibfnamefont {S.}~\bibnamefont {Qian}}, \bibinfo
  {author} {\bibfnamefont {S.}~\bibnamefont {Deng}}, \bibinfo {author}
  {\bibfnamefont {J.}~\bibnamefont {Xiao}}, \bibinfo {author} {\bibfnamefont
  {L.}~\bibnamefont {Gao}}, \bibinfo {author} {\bibfnamefont {A.~M.}\
  \bibnamefont {Levin}}, \bibinfo {author} {\bibfnamefont {Q.}~\bibnamefont
  {Li}}, \bibinfo {author} {\bibfnamefont {M.}~\bibnamefont {Lu}},\ and\
  \bibinfo {author} {\bibfnamefont {Z.}~\bibnamefont {You}},\ }\href
  {https://doi.org/10.1142/S0217751X22450014} {\bibfield  {journal} {\bibinfo
  {journal} {Int. J. Mod. Phys. A}\ }\textbf {\bibinfo {volume} {37}},\
  \bibinfo {pages} {2245001} (\bibinfo {year} {2022})},\ \Eprint
  {https://arxiv.org/abs/2204.11871} {arXiv:2204.11871 [hep-ph]} \BibitemShut
  {NoStop}%
\bibitem [{\citenamefont {Han}\ \emph {et~al.}(2021)\citenamefont {Han},
  \citenamefont {Ma},\ and\ \citenamefont {Xie}}]{Han:2020uid}%
  \BibitemOpen
  \bibfield  {author} {\bibinfo {author} {\bibfnamefont {T.}~\bibnamefont
  {Han}}, \bibinfo {author} {\bibfnamefont {Y.}~\bibnamefont {Ma}},\ and\
  \bibinfo {author} {\bibfnamefont {K.}~\bibnamefont {Xie}},\ }\href
  {https://doi.org/10.1103/PhysRevD.103.L031301} {\bibfield  {journal}
  {\bibinfo  {journal} {Phys. Rev. D}\ }\textbf {\bibinfo {volume} {103}},\
  \bibinfo {pages} {L031301} (\bibinfo {year} {2021})},\ \Eprint
  {https://arxiv.org/abs/2007.14300} {arXiv:2007.14300 [hep-ph]} \BibitemShut
  {NoStop}%
\bibitem [{\citenamefont {King}(2000)}]{King:1999kx}%
  \BibitemOpen
  \bibfield  {author} {\bibinfo {author} {\bibfnamefont {B.~J.}\ \bibnamefont
  {King}},\ }\href {https://doi.org/10.1063/1.1361674} {\bibfield  {journal}
  {\bibinfo  {journal} {AIP Conf. Proc.}\ }\textbf {\bibinfo {volume} {530}},\
  \bibinfo {pages} {142} (\bibinfo {year} {2000})},\ \Eprint
  {https://arxiv.org/abs/hep-ex/0005007} {arXiv:hep-ex/0005007} \BibitemShut
  {NoStop}%
\bibitem [{\citenamefont {Accettura}\ \emph {et~al.}(2023)\citenamefont
  {Accettura} \emph {et~al.}}]{Accettura:2023ked}%
  \BibitemOpen
  \bibfield  {author} {\bibinfo {author} {\bibfnamefont {C.}~\bibnamefont
  {Accettura}} \emph {et~al.},\ }\href
  {https://doi.org/10.1140/epjc/s10052-023-11889-x} {\bibfield  {journal}
  {\bibinfo  {journal} {Eur. Phys. J. C}\ }\textbf {\bibinfo {volume} {83}},\
  \bibinfo {pages} {864} (\bibinfo {year} {2023})},\ \Eprint
  {https://arxiv.org/abs/2303.08533} {arXiv:2303.08533 [physics.acc-ph]}
  \BibitemShut {NoStop}%
\bibitem [{\citenamefont {Abada}\ \emph {et~al.}(2019)\citenamefont {Abada}
  \emph {et~al.}}]{FCC:2018evy}%
  \BibitemOpen
  \bibfield  {author} {\bibinfo {author} {\bibfnamefont {A.}~\bibnamefont
  {Abada}} \emph {et~al.} (\bibinfo {collaboration} {FCC}),\ }\href
  {https://doi.org/10.1140/epjst/e2019-900045-4} {\bibfield  {journal}
  {\bibinfo  {journal} {Eur. Phys. J. ST}\ }\textbf {\bibinfo {volume} {228}},\
  \bibinfo {pages} {261} (\bibinfo {year} {2019})}\BibitemShut {NoStop}%
\bibitem [{\citenamefont {Bossi}\ and\ \citenamefont
  {Ciafaloni}(2020)}]{Bossi:2020yne}%
  \BibitemOpen
  \bibfield  {author} {\bibinfo {author} {\bibfnamefont {F.}~\bibnamefont
  {Bossi}}\ and\ \bibinfo {author} {\bibfnamefont {P.}~\bibnamefont
  {Ciafaloni}},\ }\href {https://doi.org/10.1007/JHEP10(2020)033} {\bibfield
  {journal} {\bibinfo  {journal} {JHEP}\ }\textbf {\bibinfo {volume} {10}},\
  \bibinfo {pages} {033}},\ \Eprint {https://arxiv.org/abs/2003.03997}
  {arXiv:2003.03997 [hep-ph]} \BibitemShut {NoStop}%
\bibitem [{Note1()}]{Note1}%
  \BibitemOpen
  \bibinfo {note}
  {Https://online.kitp.ucsb.edu/online/muoncollider-m23/}\BibitemShut {NoStop}%
\bibitem [{\citenamefont {Alwall}\ \emph {et~al.}(2014)\citenamefont {Alwall},
  \citenamefont {Frederix}, \citenamefont {Frixione}, \citenamefont {Hirschi},
  \citenamefont {Maltoni}, \citenamefont {Mattelaer}, \citenamefont {Shao},
  \citenamefont {Stelzer}, \citenamefont {Torrielli},\ and\ \citenamefont
  {Zaro}}]{Alwall:2014hca}%
  \BibitemOpen
  \bibfield  {author} {\bibinfo {author} {\bibfnamefont {J.}~\bibnamefont
  {Alwall}}, \bibinfo {author} {\bibfnamefont {R.}~\bibnamefont {Frederix}},
  \bibinfo {author} {\bibfnamefont {S.}~\bibnamefont {Frixione}}, \bibinfo
  {author} {\bibfnamefont {V.}~\bibnamefont {Hirschi}}, \bibinfo {author}
  {\bibfnamefont {F.}~\bibnamefont {Maltoni}}, \bibinfo {author} {\bibfnamefont
  {O.}~\bibnamefont {Mattelaer}}, \bibinfo {author} {\bibfnamefont {H.~S.}\
  \bibnamefont {Shao}}, \bibinfo {author} {\bibfnamefont {T.}~\bibnamefont
  {Stelzer}}, \bibinfo {author} {\bibfnamefont {P.}~\bibnamefont {Torrielli}},\
  and\ \bibinfo {author} {\bibfnamefont {M.}~\bibnamefont {Zaro}},\ }\href
  {https://doi.org/10.1007/JHEP07(2014)079} {\bibfield  {journal} {\bibinfo
  {journal} {JHEP}\ }\textbf {\bibinfo {volume} {07}},\ \bibinfo {pages}
  {079}},\ \Eprint {https://arxiv.org/abs/1405.0301} {arXiv:1405.0301 [hep-ph]}
  \BibitemShut {NoStop}%
\bibitem [{\citenamefont {Zeller}\ \emph {et~al.}(2002)\citenamefont {Zeller}
  \emph {et~al.}}]{NuTeV:2001whx}%
  \BibitemOpen
  \bibfield  {author} {\bibinfo {author} {\bibfnamefont {G.~P.}\ \bibnamefont
  {Zeller}} \emph {et~al.} (\bibinfo {collaboration} {NuTeV}),\ }\href
  {https://doi.org/10.1103/PhysRevLett.88.091802} {\bibfield  {journal}
  {\bibinfo  {journal} {Phys. Rev. Lett.}\ }\textbf {\bibinfo {volume} {88}},\
  \bibinfo {pages} {091802} (\bibinfo {year} {2002})},\ \bibinfo {note}
  {[Erratum: Phys.Rev.Lett. 90, 239902 (2003)]},\ \Eprint
  {https://arxiv.org/abs/hep-ex/0110059} {arXiv:hep-ex/0110059} \BibitemShut
  {NoStop}%
\bibitem [{\citenamefont {Davidson}\ \emph {et~al.}(2002)\citenamefont
  {Davidson}, \citenamefont {Forte}, \citenamefont {Gambino}, \citenamefont
  {Rius},\ and\ \citenamefont {Strumia}}]{Davidson:2001ji}%
  \BibitemOpen
  \bibfield  {author} {\bibinfo {author} {\bibfnamefont {S.}~\bibnamefont
  {Davidson}}, \bibinfo {author} {\bibfnamefont {S.}~\bibnamefont {Forte}},
  \bibinfo {author} {\bibfnamefont {P.}~\bibnamefont {Gambino}}, \bibinfo
  {author} {\bibfnamefont {N.}~\bibnamefont {Rius}},\ and\ \bibinfo {author}
  {\bibfnamefont {A.}~\bibnamefont {Strumia}},\ }\href
  {https://doi.org/10.1088/1126-6708/2002/02/037} {\bibfield  {journal}
  {\bibinfo  {journal} {JHEP}\ }\textbf {\bibinfo {volume} {02}},\ \bibinfo
  {pages} {037}},\ \Eprint {https://arxiv.org/abs/hep-ph/0112302}
  {arXiv:hep-ph/0112302} \BibitemShut {NoStop}%
\bibitem [{\citenamefont {Fuks}\ \emph {et~al.}(2021)\citenamefont {Fuks},
  \citenamefont {Neundorf}, \citenamefont {Peters}, \citenamefont {Ruiz},\ and\
  \citenamefont {Saimpert}}]{Fuks:2020att}%
  \BibitemOpen
  \bibfield  {author} {\bibinfo {author} {\bibfnamefont {B.}~\bibnamefont
  {Fuks}}, \bibinfo {author} {\bibfnamefont {J.}~\bibnamefont {Neundorf}},
  \bibinfo {author} {\bibfnamefont {K.}~\bibnamefont {Peters}}, \bibinfo
  {author} {\bibfnamefont {R.}~\bibnamefont {Ruiz}},\ and\ \bibinfo {author}
  {\bibfnamefont {M.}~\bibnamefont {Saimpert}},\ }\href
  {https://doi.org/10.1103/PhysRevD.103.055005} {\bibfield  {journal} {\bibinfo
   {journal} {Phys. Rev. D}\ }\textbf {\bibinfo {volume} {103}},\ \bibinfo
  {pages} {055005} (\bibinfo {year} {2021})},\ \Eprint
  {https://arxiv.org/abs/2011.02547} {arXiv:2011.02547 [hep-ph]} \BibitemShut
  {NoStop}%
\bibitem [{\citenamefont {Sj\"ostrand}\ \emph {et~al.}(2015)\citenamefont
  {Sj\"ostrand}, \citenamefont {Ask}, \citenamefont {Christiansen},
  \citenamefont {Corke}, \citenamefont {Desai}, \citenamefont {Ilten},
  \citenamefont {Mrenna}, \citenamefont {Prestel}, \citenamefont {Rasmussen},\
  and\ \citenamefont {Skands}}]{Sjostrand:2014zea}%
  \BibitemOpen
  \bibfield  {author} {\bibinfo {author} {\bibfnamefont {T.}~\bibnamefont
  {Sj\"ostrand}}, \bibinfo {author} {\bibfnamefont {S.}~\bibnamefont {Ask}},
  \bibinfo {author} {\bibfnamefont {J.~R.}\ \bibnamefont {Christiansen}},
  \bibinfo {author} {\bibfnamefont {R.}~\bibnamefont {Corke}}, \bibinfo
  {author} {\bibfnamefont {N.}~\bibnamefont {Desai}}, \bibinfo {author}
  {\bibfnamefont {P.}~\bibnamefont {Ilten}}, \bibinfo {author} {\bibfnamefont
  {S.}~\bibnamefont {Mrenna}}, \bibinfo {author} {\bibfnamefont
  {S.}~\bibnamefont {Prestel}}, \bibinfo {author} {\bibfnamefont {C.~O.}\
  \bibnamefont {Rasmussen}},\ and\ \bibinfo {author} {\bibfnamefont {P.~Z.}\
  \bibnamefont {Skands}},\ }\href {https://doi.org/10.1016/j.cpc.2015.01.024}
  {\bibfield  {journal} {\bibinfo  {journal} {Comput. Phys. Commun.}\ }\textbf
  {\bibinfo {volume} {191}},\ \bibinfo {pages} {159} (\bibinfo {year}
  {2015})},\ \Eprint {https://arxiv.org/abs/1410.3012} {arXiv:1410.3012
  [hep-ph]} \BibitemShut {NoStop}%
\bibitem [{\citenamefont {Cacciari}\ \emph {et~al.}(2012)\citenamefont
  {Cacciari}, \citenamefont {Salam},\ and\ \citenamefont
  {Soyez}}]{Cacciari:2011ma}%
  \BibitemOpen
  \bibfield  {author} {\bibinfo {author} {\bibfnamefont {M.}~\bibnamefont
  {Cacciari}}, \bibinfo {author} {\bibfnamefont {G.~P.}\ \bibnamefont
  {Salam}},\ and\ \bibinfo {author} {\bibfnamefont {G.}~\bibnamefont {Soyez}},\
  }\href {https://doi.org/10.1140/epjc/s10052-012-1896-2} {\bibfield  {journal}
  {\bibinfo  {journal} {Eur. Phys. J. C}\ }\textbf {\bibinfo {volume} {72}},\
  \bibinfo {pages} {1896} (\bibinfo {year} {2012})},\ \Eprint
  {https://arxiv.org/abs/1111.6097} {arXiv:1111.6097 [hep-ph]} \BibitemShut
  {NoStop}%
\bibitem [{\citenamefont {Boronat}\ \emph {et~al.}(2018)\citenamefont
  {Boronat}, \citenamefont {Fuster}, \citenamefont {Garcia}, \citenamefont
  {Roloff}, \citenamefont {Simoniello},\ and\ \citenamefont
  {Vos}}]{Boronat_2018}%
  \BibitemOpen
  \bibfield  {author} {\bibinfo {author} {\bibfnamefont {M.}~\bibnamefont
  {Boronat}}, \bibinfo {author} {\bibfnamefont {J.}~\bibnamefont {Fuster}},
  \bibinfo {author} {\bibfnamefont {I.}~\bibnamefont {Garcia}}, \bibinfo
  {author} {\bibfnamefont {P.}~\bibnamefont {Roloff}}, \bibinfo {author}
  {\bibfnamefont {R.}~\bibnamefont {Simoniello}},\ and\ \bibinfo {author}
  {\bibfnamefont {M.}~\bibnamefont {Vos}},\ }\bibfield  {journal} {\bibinfo
  {journal} {The European Physical Journal C}\ }\textbf {\bibinfo {volume}
  {78}},\ \href {https://doi.org/10.1140/epjc/s10052-018-5594-6}
  {10.1140/epjc/s10052-018-5594-6} (\bibinfo {year} {2018})\BibitemShut
  {NoStop}%
\bibitem [{\citenamefont {de~Favereau}\ \emph {et~al.}(2014)\citenamefont
  {de~Favereau}, \citenamefont {Delaere}, \citenamefont {Demin}, \citenamefont
  {Giammanco}, \citenamefont {Lema\^\i{}tre}, \citenamefont {Mertens},\ and\
  \citenamefont {Selvaggi}}]{deFavereau:2013fsa}%
  \BibitemOpen
  \bibfield  {author} {\bibinfo {author} {\bibfnamefont {J.}~\bibnamefont
  {de~Favereau}}, \bibinfo {author} {\bibfnamefont {C.}~\bibnamefont
  {Delaere}}, \bibinfo {author} {\bibfnamefont {P.}~\bibnamefont {Demin}},
  \bibinfo {author} {\bibfnamefont {A.}~\bibnamefont {Giammanco}}, \bibinfo
  {author} {\bibfnamefont {V.}~\bibnamefont {Lema\^\i{}tre}}, \bibinfo {author}
  {\bibfnamefont {A.}~\bibnamefont {Mertens}},\ and\ \bibinfo {author}
  {\bibfnamefont {M.}~\bibnamefont {Selvaggi}} (\bibinfo {collaboration}
  {DELPHES 3}),\ }\href {https://doi.org/10.1007/JHEP02(2014)057} {\bibfield
  {journal} {\bibinfo  {journal} {JHEP}\ }\textbf {\bibinfo {volume} {02}},\
  \bibinfo {pages} {057}},\ \Eprint {https://arxiv.org/abs/1307.6346}
  {arXiv:1307.6346 [hep-ex]} \BibitemShut {NoStop}%
\bibitem [{muc()}]{mucard}%
  \BibitemOpen
  \href@noop {} {\bibinfo {title} {Muon collider delphes card}},\ \bibinfo
  {howpublished}
  {\url{https://github.com/delphes/delphes/blob/master/cards/delphes\_card\_MuonColliderDet.tcl}}\BibitemShut
  {NoStop}%
\bibitem [{\citenamefont {Yang}\ \emph {et~al.}(2021)\citenamefont {Yang},
  \citenamefont {Qian}, \citenamefont {Guan}, \citenamefont {Li}, \citenamefont
  {Meng}, \citenamefont {Xiao}, \citenamefont {Lu},\ and\ \citenamefont
  {Li}}]{Yang:2021zak}%
  \BibitemOpen
  \bibfield  {author} {\bibinfo {author} {\bibfnamefont {T.}~\bibnamefont
  {Yang}}, \bibinfo {author} {\bibfnamefont {S.}~\bibnamefont {Qian}}, \bibinfo
  {author} {\bibfnamefont {Z.}~\bibnamefont {Guan}}, \bibinfo {author}
  {\bibfnamefont {C.}~\bibnamefont {Li}}, \bibinfo {author} {\bibfnamefont
  {F.}~\bibnamefont {Meng}}, \bibinfo {author} {\bibfnamefont {J.}~\bibnamefont
  {Xiao}}, \bibinfo {author} {\bibfnamefont {M.}~\bibnamefont {Lu}},\ and\
  \bibinfo {author} {\bibfnamefont {Q.}~\bibnamefont {Li}},\ }\href
  {https://doi.org/10.1103/PhysRevD.104.093003} {\bibfield  {journal} {\bibinfo
   {journal} {Phys. Rev. D}\ }\textbf {\bibinfo {volume} {104}},\ \bibinfo
  {pages} {093003} (\bibinfo {year} {2021})},\ \Eprint
  {https://arxiv.org/abs/2107.13581} {arXiv:2107.13581 [hep-ph]} \BibitemShut
  {NoStop}%
\bibitem [{\citenamefont {Tumasyan}\ \emph {et~al.}(2023)\citenamefont
  {Tumasyan} \emph {et~al.}}]{CMS:2022hvh}%
  \BibitemOpen
  \bibfield  {author} {\bibinfo {author} {\bibfnamefont {A.}~\bibnamefont
  {Tumasyan}} \emph {et~al.} (\bibinfo {collaboration} {CMS}),\ }\href
  {https://doi.org/10.1103/PhysRevLett.131.011803} {\bibfield  {journal}
  {\bibinfo  {journal} {Phys. Rev. Lett.}\ }\textbf {\bibinfo {volume} {131}},\
  \bibinfo {pages} {011803} (\bibinfo {year} {2023})},\ \Eprint
  {https://arxiv.org/abs/2206.08956} {arXiv:2206.08956 [hep-ex]} \BibitemShut
  {NoStop}%
\end{thebibliography}%
\end{document}